\documentclass[11pt]{article}
\usepackage{epsfig}
\def\beq{\begin{equation}} \def\eeq{\end{equation}}
\def\beqa{\begin{eqnarray}} \def\eeqa{\end{eqnarray}}
\def\bce{\begin{center}}  \def\ece{\end{center}}
\def\bfig{\begin{figure}}  \def\efig{\end{figure}}
\def\bit{\begin{itemize}}    \def\eit{\end{itemize}}
\def\ben{\begin{enumerate}}    \def\een{\end{enumerate}}
\def\xs{cross-section} \def\xss{cross-sections}
\def\npb{Nucl. Phys. {\bf B}} \def\plb{Phys. Lett. {\bf B}}
\def\prd{Phys. Rev. {\bf D}}
 
\def\cl{{\cal L}}

\begin{document}

\rightline{FNT/T-99/18}

\vspace{0.5truecm}

\begin{center}
{\Large \bf Electroweak physics in six-fermion final states}\\
{\Large \bf at future $e^+ e^-$ colliders}
\end{center}

\noindent
\begin{center}
F.~Gangemi$^{1,2}$, G.~Montagna$^{1,2}$, M.~Moretti$^{3}$, \\
O.~Nicrosini$^{2,1}$  and F.~Piccinini$^{2,1}$
\end{center}

\vskip 0.2cm
\noindent
\begin{center}
$^1$ Dipartimento di Fisica Nucleare e Teorica, Universit\`a di Pavia,
Pavia, Italy\\
$^2$ Istituto Nazionale di Fisica Nucleare, Sezione di Pavia, Pavia, Italy\\
$^3$ Dipartimento di Fisica, Universit\`a di Ferrara and INFN, Sezione di
Ferrara, Ferrara, Italy
\end{center}

\vskip 0.3cm

\begin{abstract}
Three studies on six-fermion production processes are presented, in which the
production of an intermediate-mass Higgs boson, the {\it top}-quark physics and
the analysis of possible anomalous quartic gauge couplings are considered.
A Monte Carlo event generator has been developed for full electroweak
tree-level calculations on six-fermion processes at the Linear Collider. The
Monte Carlo procedure has been adapted to deal with a large variety of diagram
topologies, so as to keep under control the relevant signals and all the
backgrounds of such processes. The effects of initial-state-radiation and
beamstrahlung are also taken into account. The relevance of electroweak
backgrounds and finite-width effects is discussed and several analyses of
final-state distributions are presented, both for the detection of the signals
of interest and for the study of properties of the particles involved, thus
showing the importance of complete calculations for precision studies at the
Linear Collider.

\vskip 0.2cm

\end{abstract}

\section{Introduction}
\label{sec:intro}
Many signals of interest for tests of the Standard Model and search for new
physics at the Linear Collider will be given by many-particle final states. It
is therefore important to develop the calculation techniques and the tools
necessary for the physics analysis of these phenomena, taking into account all
the background effects and keeping under control all the relevant final-state
correlations.

In particular, the six-fermion signatures will be relevant to several subjects,
such as intermediate-mass Higgs boson production, {\it top}-quark physics and
the analysis of anomalous quartic gauge couplings. These topics are addressed
in the present contribution. Numerical results are presented and discussed.

The numerical calculations have been performed by means of a computer code
that involves the algorithm ALPHA \cite{alpha}, for the automatic calculation
of the scattering amplitudes, and a Monte Carlo integration procedure derived
from the four-fermion codes HIGGSPV \cite{higgspv} and WWGENPV \cite{wwgenpv},
and developed to deal with six-fermion processes.

\section{Intermediate-mass Higgs boson}
\label{sec:Higgs}

The search for the Higgs boson, that is carried on presently at LEP and
Tevatron, will be also in the physics programme of future high-energy colliders,
where the whole range of mass values allowed by the general consistency
conditions for the Standard Model, that is up to $\simeq 1$ TeV, can be
explored.

The current lower bound on the Higgs mass deduced from direct search at LEP is
95.2 GeV at 95 $\%$ C.L. \cite{tamp99}, while the upper bound given by fits to
the precision data on electroweak observables is 245 GeV at 95 $\%$ C.L.
\cite{tamp99}.

The Linear Collider will not only be able to discover the Higgs boson, but it
will also provide the possibility of making precision studies on its properties.
It is then of great interest to make accurate predictions on the processes in
which the Higgs boson can be produced at the LC, and to develop the tools for
making simulations. In the mass range favoured by the present experimental
information, that is between 100 and 250 GeV, the relevant signatures are
four-fermion final states if the Higgs mass is below 130-140 GeV, and
six-fermion final states if the Higgs mass is greater than 140 GeV. The
processes of the first kind have been extensively studied in connection with
physics at LEP, while those of the second kind have only recently been addressed
\cite{sixfzpc}--\cite{6fhiggs}.

In this contribution, complete electroweak tree-level calculations for the
processes $e^+e^-\to q\overline q l^+ l^- \nu\overline\nu$, with $q=u,d,c,s$,
$l=e,\mu,\tau$ and $\nu=\nu_e,\nu_\mu,\nu_\tau$ are presented. These processes
are characterized by the presence of both charged and neutral currents and of
different mechanisms of Higgs production involving Higgs-strahlung and vector
boson fusion; moreover, QCD backgrounds are absent.

The total cross-section is shown in fig.~\ref{fig:6fsscan20} as a function of
the center-of-mass (c.m.) energy for three values of the Higgs mass, with
suitable kinematical cuts, to avoid the soft-pair singularities.
The off-shellness effects due to the finite widths of the gauge bosons and of
the Higgs boson have also been studied by comparing the result obtained by
means of the signal diagrams with the one in the narrow-width approximation.
Deviations of the order of $10-15\%$ have been found \cite{6fhiggs}.

Various distributions have been studied, after generating samples of
unweighted events. The analysis is restricted in this case to the processes
with $l=e$ and a luminosity of 500 fb$^{-1}$ is assumed. The invariant masses
of different systems of four fermions are plotted in fig.~\ref{fig:nt185ibs4},
including the effects of initial-state radiation (ISR) \cite{sf} and
beamstrahlung (BS) \cite{circe}. The different sets of fermions correspond to
the Higgs boson in different signal diagrams. It is interesting to observe that
at 800 GeV the $qqe^+e^-$ invariant mass gives a clean signal, not affected by
ISR and BS, that can be traced back to the $WW$ fusion signal diagram.

\bfig
\bce
\epsfig{file=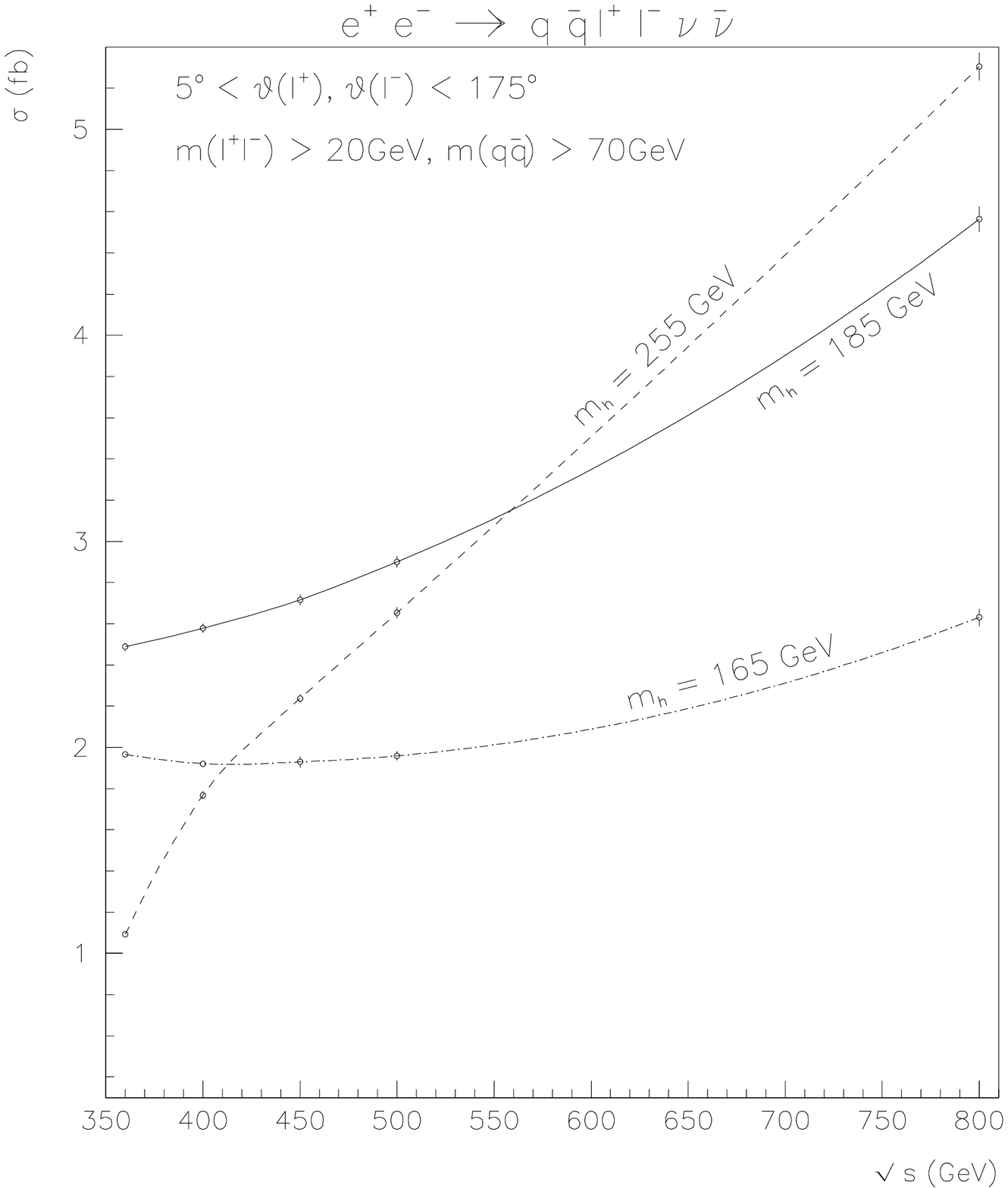,height=7.8cm,width=7.8cm}
\caption{\small Total cross section for the process 
 $e^+e^-\to q\overline ql^+l^-\nu\overline\nu$ in the Born approximation,
 as a function of $\sqrt s$ for three different values of the Higgs mass
$m_H$. The angles $\theta(l^+)$, $\theta(l^-)$
of the charged leptons with the beam axis are in the interval 
$5^\circ$-$175^\circ$,
the $e^+e^-$ and the $q\bar q$ invariant masses are larger than $20$ GeV.}
\label{fig:6fsscan20}
\ece
\efig
Other distributions can be considered in order to reveal the presence of the
Higgs boson and to measure its properties \cite{6fhiggs}.
As a conclusion, the processes under consideration turn out to be of interest
for the study of intermediate Higgs bosons. Thanks to the sums over quark,
charged lepton and neutrino flavours, as well as the combined action of
different production mechanisms, assuming a luminosity of $500$ fb$^{-1}$/yr
and a Higgs mass of, say, $185$ GeV, more than $1000$ events can be expected
at a c.m. energy of $360$ GeV and more than $2000$ at $800$ GeV
(see fig.~\ref{fig:6fsscan20}). The complete calculation shows the relevance
of background and off-shellness effects, and it is possible to exploit the
features of the different signal diagrams to find at different energies
suitable distributions that are sensitive to the presence and to the properties
of the Higgs boson.
\bfig
\bce
\epsfig{file=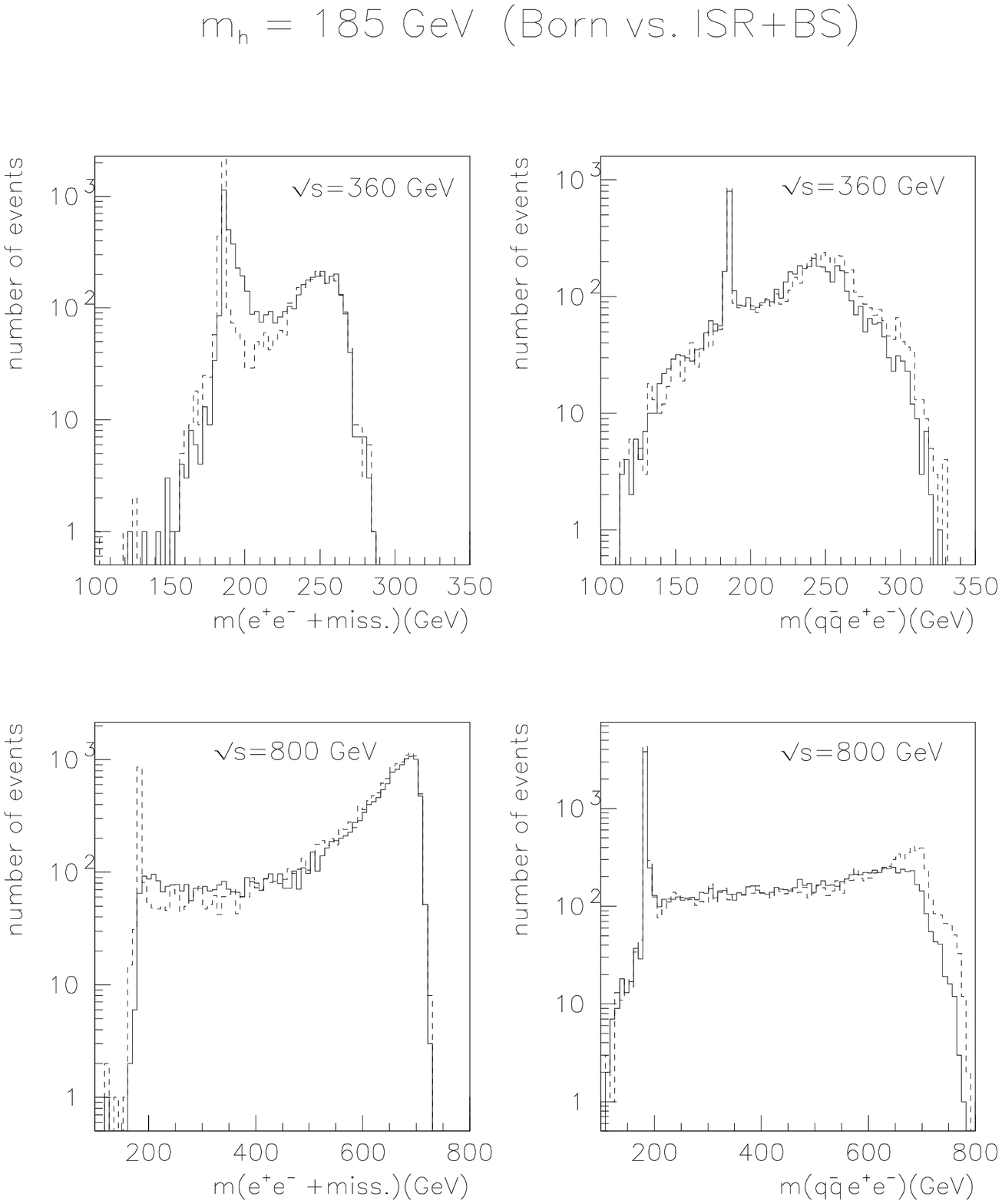,height=8.8cm,width=8.8cm}
\caption{\small Invariant-mass distributions for four-fermion systems in the
Born approximation (dashed histograms) and with ISR and beamstrahlung (solid
histograms) at $\sqrt s=360$ GeV (upper row) and $\sqrt s=800$ GeV
(lower row).}
\label{fig:nt185ibs4}
\ece
\efig
\section{\bf {\it Top}-quark physics in six-quark processes}
\label{sec:top}

The study of $t\bar t$ production both at threshold and above at the
Linear Collider will give the opportunity of making significant tests of QCD
and to get important information through the determination of the electroweak
properties of the {\it top} quark.

The production of a $t\bar t$ pair gives rise to six fermions in the
final state. The $6f$ signatures relevant to the study of the {\it top} quark
in $e^+e^-$ collisions can be summarized as follows: $b\bar bl\nu_ll'\nu_{l'}$
(leptonic, $\sim 10\%$ of the total rate), $b\bar bq\bar q'l\nu_l$ (semi
leptonic, $\sim 45\%$), $b\bar b+4q\quad$ (hadronic, $\sim 45\%$).
Semi leptonic signatures have been considered in refs.~\cite{to1,kek,4jln}.
It is then of great interest to carefully evaluate the size of the totally
hadronic, six-quark ($6q$) contributions to integrated cross-sections and
distributions as well as to determine their phenomenological features.

The $6q$ signatures of the form $b\bar b+4q$, where $q=u,d,c,s$ are considered
in the present study and the results of complete electroweak tree-level
calculations are presented. In particular the r\^ole of electroweak backgrounds
and of ISR and BS are studied and the shape of the events is analysed to the
end of isolating the QCD backgrounds.

The integrated \xs\ has been studied in the energy range between 350 and 800 GeV
for different Higgs masses and has been compared with the signal contribution
alone and with the result in the narrow-width-approximation, showing that the
background and off-shellness effects are of the order of several per cent
\cite{6ftop}.

The total electroweak \xs\ has also been studied at the threshold for $t\bar t$
production as a function of the Higgs mass. Although the dominant effects in
this case come from QCD contributions, as is well known \cite{teub}, the
electroweak backgrounds turn out to give a sizeable uncertainty, of the order
of $10\%$ of the pure electroweak contribution, in the intermediate range of
Higgs masses (see fig.~\ref{fig:hmscan}), which is related to the fact that the
Higgs mass is not known.
\bfig
\bce
\epsfig{file=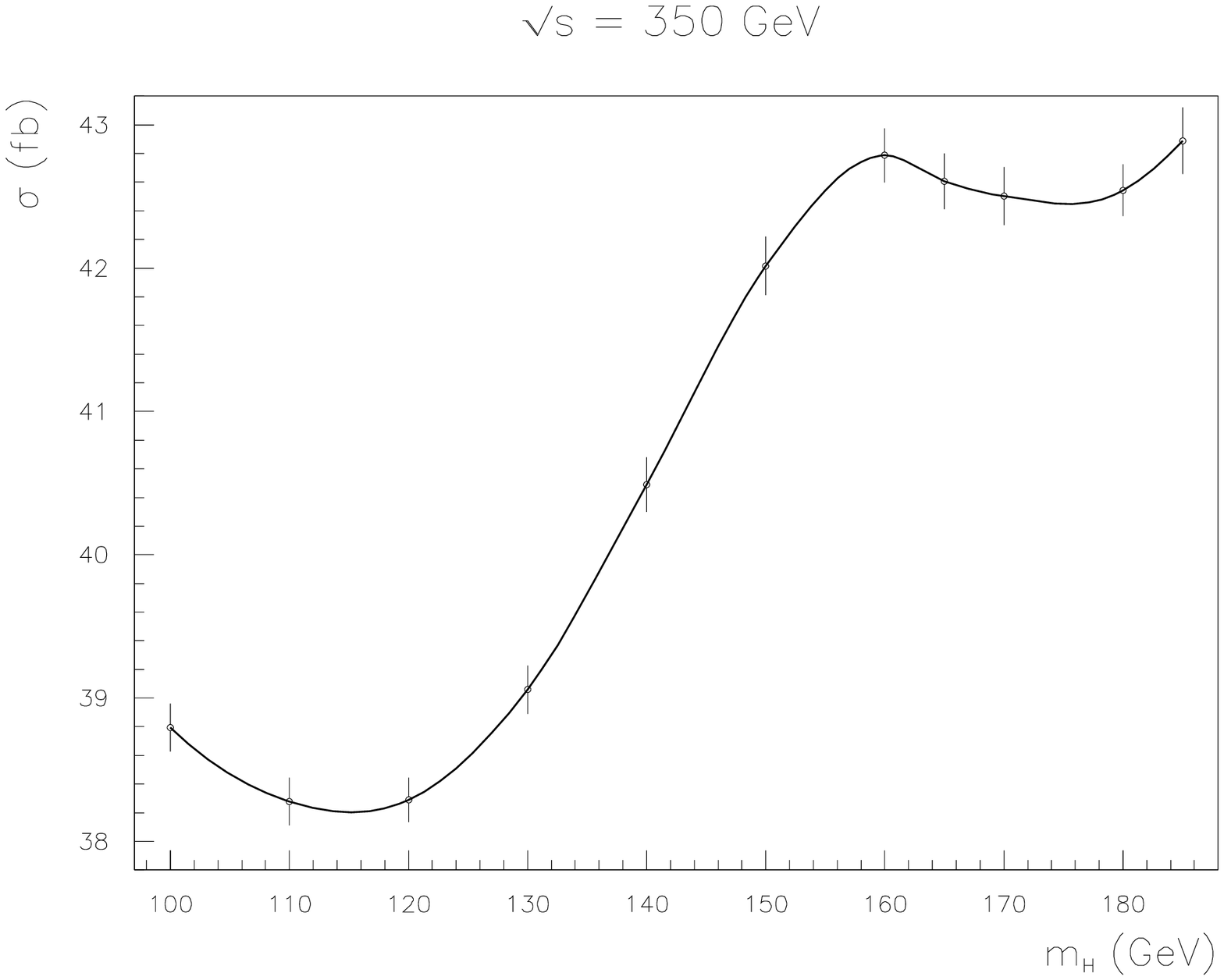,height=6.6cm,width=8.8cm}
\caption{\small Total cross-section as a function of the Higgs mass at the
threshold for $t\bar t$ production.}
\label{fig:hmscan}
\ece
\efig
The topology of the events has been studied by means of various event-shape
variables, in order to study the possibility of isolating the {\it top}-quark
signal from the QCD backgrounds. The pure QCD contributions have been analysed
in ref.~\cite{sixj}. In fig.~\ref{fig:shape} the thrust distribution of the
electroweak contribution is shown at a c.m. energy of 500 GeV and with a Higgs
mass of 185 GeV in the Born approximation (dashed histogram) and with ISR and
BS (the solid histogram; in this case, the distribution is calculated after
going to the c.m. frame). A luminosity of 500 fb$^{-1}$ is assumed and the
invariant masses of the $b\bar b$ pair and of all the pairs of quarks other
than $b$ are required to be greater than 10 GeV. Remarkable effects due to ISR
and BS can be seen in this plot, where the peak in the thrust distribution is
strongly reduced with respect to the Born approximation and the events are
shifted towards the lower values of $T$, which correspond to spherical events.
As a conclusion, at 500 GeV, in view of the results of ref.~\cite{sixj}, the
thrust variable is very effective in discriminating pure QCD backgrounds, also
in the presence of electroweak backgrounds and of ISR and BS.
\bfig
\bce
\epsfig{file=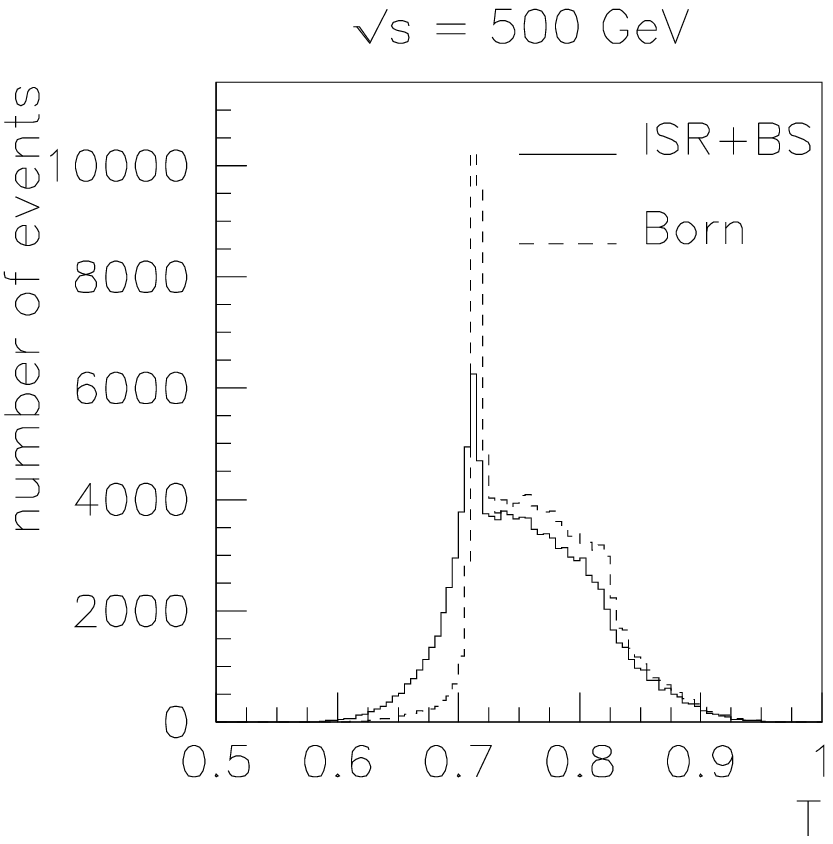,height=7.cm,width=7.cm}
\caption{\small Thrust distribution in the Born approximation (dashed histogram)
and with initial-state radiation and beamstrahlung (solid histogram), at a
c.m. energy of 500 GeV and for a Higgs mass of 185 GeV. A luminosity of 500
fb$^{-1}$ is assumed. The invariant masses of the $b\bar b$ pair and of all the
pairs of quarks other than $b$ are required to be greater than 10 GeV.}
\label{fig:shape}
\ece
\efig
\section{Anomalous gauge couplings}
\label{sec:agc}

The situation in which a Higgs boson with mass below 1 TeV is not found can be
described by means of the electroweak effective lagrangian, as discussed in
ref.~\cite{ewchl}. Different models of electroweak symmetry breaking can be
parameterized by this effective lagrangian. The contributions at lowest order
in the chiral expansion, if the $SU(2)$ custodial symmetry is assumed, are
model-independent. At next-to-leading order, dimension-four operators are
present, with parameters that depend on the model of symmetry breaking adopted.
These operators can give rise to trilinear and quadrilinear couplings of the
massive gauge bosons, that modify the standard ones contained in the Yang-Mills
lagrangian for the gauge bosons. The dimension-four operators that give only
quartic vertices, usually indicated as $\cl_4$, $\cl_5$, $\cl_6$, $\cl_7$ and
$\cl_{10}$, have been implemented in ALPHA. Anomalous $4W$, $WWZZ$ and $4Z$
vertices are provided by these terms. In the following only the two
$SU(2)$-custodial symmetry conserving operators $\cl_4$ and $\cl_5$ are
discussed. Their expressions in the unitary gauge are \cite{ewchl}:
\beqa
\nonumber
\label{eq:agc1}
\cl_4&=&
\alpha_4g^4\left(\frac{1}{2}W^+_\mu W^{+\mu}W^-_\nu W^{-\nu}
+\frac{1}{2}(W^+_\mu W^{-\mu})^2+\frac{1}{c_W^2}W^+_\mu Z^\mu W^-_\nu Z^\nu+
\frac{1}{4c_W^4}(Z_\mu Z^\mu)^2\right)\\
\label{eq:agc2}
\cl_5&=&
\alpha_5\left((W^+_\mu W^{-\mu})^2+\frac{1}{c_W^2}W^+_\mu W^{-\mu} Z_\nu Z^\nu
+\frac{1}{4c_W^4}(Z_\mu Z^\mu)^2\right)
\eeqa

These anomalous quartic couplings have been studied by several authors at the
loop level, where they contribute to radiative corrections to electroweak
observables~\cite{agcloop}, and at tree-level in processes of gauge boson
scattering and real gauge boson production \cite{qgcm}. In a more realistic
approach, where the gauge bosons are not real, signatures with at least six
fermions in the final state have to be considered. For the present study,
the processes $e^+e^-\to 2q+2q'\nu_e\bar\nu_e$, with $q=u,c$ and $q'=d,s$, have
been considered, and a full tree-level calculation has been performed, by using
the effective lagrangian containing the dimension-four operators mentioned
above.

Through the study of some event samples, several variables that are sensitive
to the parameters $\alpha_4$ and $\alpha_5$ have been found.
A set of kinematical cuts has thus been deduced to enhance the effects of
anomalous couplings. For example, the \xs\ obtained with this set of cuts is
shown in fig.~\ref{fig:a45fine} as a function of the parameter $\alpha_5$ in
the range $(-0.01,0.01)$. The variables involved in the cuts, as indicated in
the figure, are the invariant mass of the system of four jets, $M(WW)$,
the angle $\theta(W)$ of one reconstructed $W$ (where a simple procedure is used
to identify the $W$ boson from the quarks) and the invariant mass of the pair
of jets with lowest transverse momentum. The limits of the $1\sigma$
experimental uncertainty around the value at $\alpha_5=0$ are also shown in
fig.~\ref{fig:a45fine}, by assuming a luminosity of 1000 fb$^{-1}$. The
sensitivity to this parameter with this set of cuts can be seen to be of the
order of $10^{-2}$.
The \xss\ and the variables used in the cuts have been analysed also in the 
presence of ISR and BS. The above conclusions on the sensitivity to the
anomalous couplings should not be modified by the inclusion of such effects
as long as variables not involving the missing momentum are considered.
\bfig
\bce
\epsfig{file=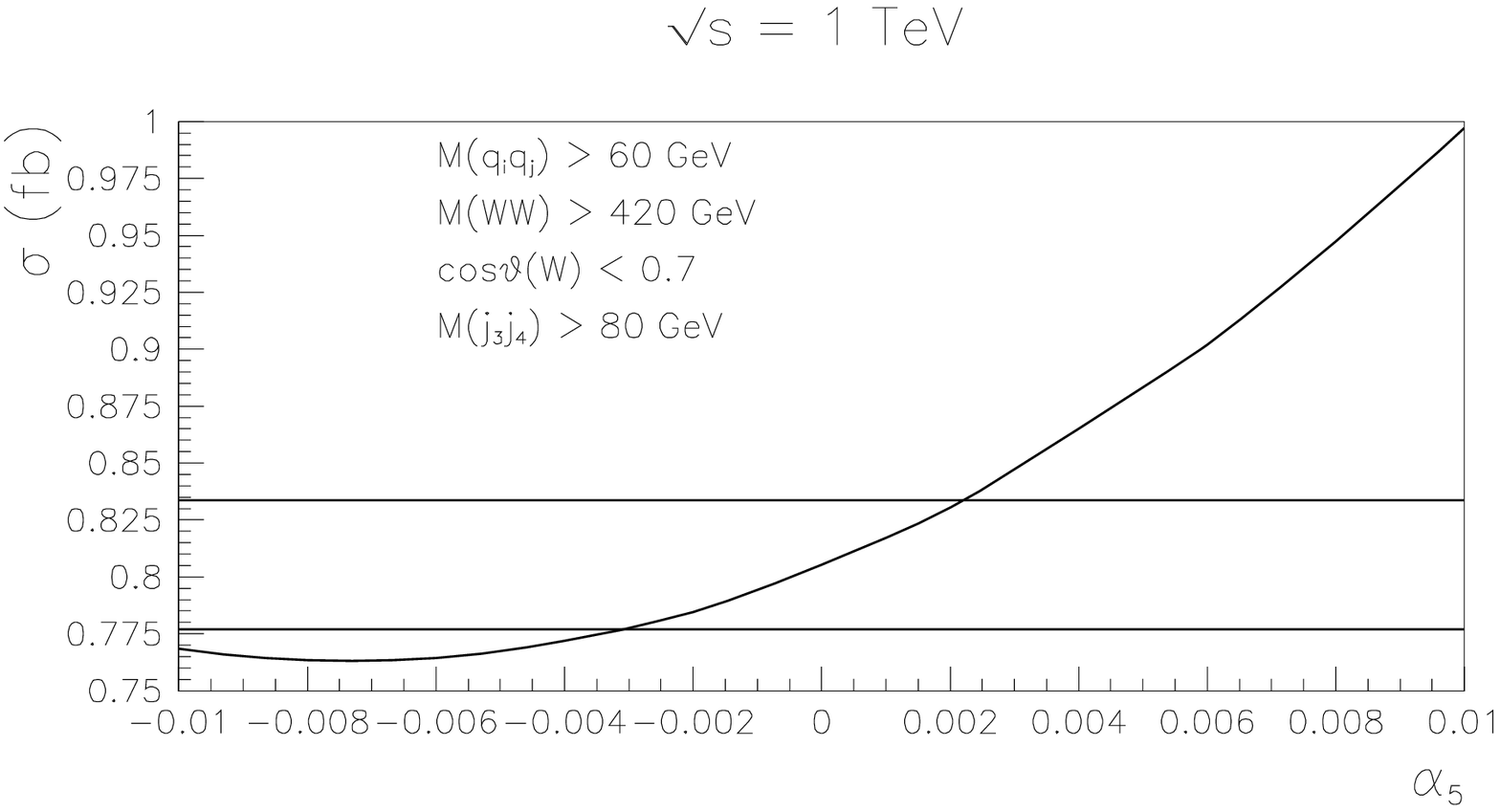,height=6.cm,width=11.cm}
\caption{\small Integrated \xs\ in the Born approximation at a c.m. energy of
 1 TeV, as a function of $\alpha_5$ with a set of kinematical cuts studied to
 enhance the sensitivity to the anomalous couplings. The horizontal lines
 are the $1 \sigma$ bounds around the $\alpha_5=0$ value, with the experimental
 uncertainty corresponding to a luminosity of 1000 fb$^{-1}$.}
\label{fig:a45fine}
\ece
\efig
\section{Conclusions}
\label{sec:concl}
The six-fermion final states will be among the most relevant new signatures at
future $e^+e^-$ linear colliders. In particular they are interesting for
Higgs bosons in the intermediate mass range, for $t\bar t$ production and for
the study of quartic anomalous gauge couplings. These subjects are addressed in
the studies that are presented in this contribution.

A Monte Carlo event generator has been developed for complete tree-level
calculations of such processes at the energies of the Linear Collider. This
code, that makes use of ALPHA for the calculation of the scattering amplitudes,
has been adapted to deal with a large variety of diagram topologies, including
both charged and neutral currents, so as to keep under control all the relevant
signals of interest as well as the complicated backgrounds that are involved
in six-fermion processes where hundreds of diagrams contribute to the
tree-level amplitudes. The effects of initial-state-radiation and
beamstrahlung are also included.

The studies of Higgs boson production in the intermediate mass range, of
$t\bar t$ production and of anomalous gauge couplings have shown the importance
of complete calculations to keep under control all the background and
finite-width effects, and to obtain simulations of all kinds of final-state
distributions such as invariant masses, angular correlations and event-shape
variables, that are essential both for the detection of the signals of interest
and for the analysis of the properties of the particles under study.

\vspace{0.4truecm}
{\bf Acknowledgements}\\
The authors wish to thank Thorsten Ohl for his interest in this work and for
useful discussions.

\end{document}